\documentclass[useAMS,usenatbib,usegraphics]{mn2e}
\usepackage{graphicx,epsfig}  
\usepackage{amsmath,amssymb}
\def\ut#1{\mathop{\vtop{\ialign{##\crcr
     $\hfil\displaystyle{#1}\hfil$\crcr\noalign
     {\kern1pt\nointerlineskip}\hbox{$\hfil\sim\hfil$}\crcr
     \noalign{\kern1pt}}}}}

\def\undersymbol#1#2{\mathop{\vtop{\ialign{##\crcr
     $\hfil\displaystyle{#2}\hfil$\crcr\noalign
     {\kern1pt\nointerlineskip}\hbox{$\hfil#1\hfil$}\crcr
     \noalign{\kern1pt}}}}}

\title[Measuring Polarization in microlensing events]
{Measuring polarization in microlensing events}
\author[G. Ingrosso et al.]
{G. Ingrosso,$^{1,2}$
\thanks{E-mail: ingrosso@le.infn.it}
 S. Calchi Novati,$^{3,4,5}$
 F. De Paolis,$^{1,2}$
\newauthor Ph. Jetzer,$^6$
A. A. Nucita,$^{1,2}$ and 
F. Strafella$^{1,2}$ \\
$^1$  Dipartimento di Matematica e Fisica, ``Ennio De Giorgi'', 
Universit\`a del Salento, CP 193, I-73100 Lecce, Italy \\
$^2$   {\it INFN} Sezione di Lecce, CP 193, I-73100 Lecce, Italy \\
$^3$  Dipartimento di Fisica ''E. R. Caianiello'', 
Universit\'a degli Studi di Salerno, 
Via Giovanni Paolo II, 132, 84084 Fisciano (SA) - Italy \\
$4$ Istituto Internazionale per gli Alti Studi Scientifici (IIASS), 
Vietri Sul Mare (SA), Italy\\
$5$ NASA Exoplanet Science Institute, MS 100-22, California Institute of Technology, 
Pasadena CA 91125, USA (Sagan visiting fellow) \\
$^6$       Physics Department,
           University of  Z\"{u}rich, Winterthurerstrasse 190,
           CH-8057 Z\"{u}rich, Switzerland }
\begin{document}
\date{Accepted xxx; Received xxx;
in original form xxx}
\pagerange{\pageref{firstpage}--\pageref{lastpage}} \pubyear{2011}
\maketitle
\label{firstpage}
\begin{abstract}
 
We re-consider the polarization of the star light that may arise during  microlensing
events due to the high gradient of magnification across the atmosphere of the source star,
by exploring the full range of microlensing and stellar physical parameters.
Since it is already known that
only cool evolved giant stars give rise to the highest polarization signals, we follow 
the model by \cite{Simmons02} to compute the polarization as due to the photon scattering 
on dust grains in the stellar {wind}. 

Motivated by the possibility to perform a polarization measurement during an ongoing
microlensing event, we consider the recently reported event catalog by the OGLE 
collaboration covering the 2001-2009 campaigns (OGLE-III events), that makes available 
the largest and more comprehensive set of single lens microlensing events towards the
Galactic bulge. 

The study of these events, integrated by a Monte Carlo analysis, allows us to estimate the
expected polarization profiles and to predict 
for which source stars and at which time is most convenient
to perform a polarization measurement in an ongoing event.

We find that about two dozens of OGLE-III events (about 1 percent of the total)
have maximum polarization degree in the range $0.1 < P_{\rm max} <1$ percent, 
corresponding to source stars with apparent magnitude $I  \ut < 14.5$,
being very cool red giants.
This signal is measurable   
by using the  FORS2 polarimeter at VLT telescope with about 1 hour integration time. 

\end{abstract}

\section{Introduction}

It is well known that the light received from the stars is 
linearly polarized by the photon scattering occurring in the 
stellar atmospheres. The mechanism is particularly effective 
for the hot stars that have a free electron atmosphere, giving
rise to a maximum polarization degree that can be as high as 
about 12 percent at the stellar limb \citep{Chandra60}. 
By a minor extent, 
polarization may be also induced in main sequence stars (of late type)
by the scattering of star light off atoms and molecules \citep{Stenflo99} 
and in evolved, cool giant stars by photon scattering on dust grains contained 
in their extended envelopes powered by strong stellar winds
\citep{Simmons95a}.

The idea that the polarization could be enhanced by the microlensing 
effect at a level eventually above the instrumental threshold 
was first investigated in relation
to supernovae by \cite{SW87}. 
Then, \cite{Simmons95a}, \cite{Simmons95b} and \cite{BCS96} presented a numerical
calculation of the polarization induced by a single-lens (the Schwarzschild lens)
microlensing a hot source star.
The maximum polarization degree was found to be about $0.1$ per cent. 
Hence, \cite{Agol96} calculated the time-dependent 
polarization of a hot star being microlensed by a binary system and found that 
polarization signals as high as $P \simeq 1$ {percent} can be achieved when the 
source crosses in the lens plane a caustic \citep{SEF}
or passes near a cusp \citep{Schneider92,Zakharov95}.

From an observational perspective the more promising
line of sight for detecting polarization signals
associated to microlensing events is that
towards the Galactic bulge where by now each year
of order one thousand new events are reported,
with the primary scientific goal being
the search and characterization of exoplanets \citep{Gaudi10,Dominik10}.
Along that line of sight, however, we expect a
few or none hot giant source stars. Rather,
it is expected that the highest 
polarization signals occur for microlensed cool, giant stars 
endowed with extented dust grain envelopes \citep{Ingrosso12}.
Indeed, these evolved stars 
constitute a significant fraction of 
the lensed sources in events towards the Galactic bulge, the LMC \citep{Alcock97,Moniez10} 
and the M31 galaxy \citep{Sebastiano10,Sebastiano14}.

A useful formalism to evaluate the polarization profiles towards microlensed
cool giant source stars has been developed by \cite{Simmons02} for single lens events and 
by \cite{Ignace06} for binary events. It turned out that the polarization degree ultimately 
depends on the dust optical depth $\tau$ and that for large enough values of 
$\tau \simeq 10^{-1}$ the polarization degree might become close to that expected
for hot giant stars. 

In a recent work \citep{Ingrosso12} a specific set 
of highly magnified, single-lens events \citep{Choi} was considered along with a subset 
of exoplanetary events observed towards the Galactic bulge \citep{Gaudi10,Dominik10}. 
For these events the polarization profiles as a function of time were calculated, 
taking into account the nature of the source stars
(a main sequence of late type or a cool giant star) and it was shown that 
the currently available technology may potentially allow the detection of such signals.

Besides the interest related to stellar astrophysics, the observation of the polarization 
variability may in principle provide independent constraints on the lensing parameters. 
Indeed, it was shown \citep{Ingrosso14} that polarization 
measurementes may allow to distinguish between binary and exoplanetary 
\footnote{Exoplanetary events are binary lens systems characterized by values
of the planet-to-star mass ratio $q \ll 1$ and smaller star-to-planet 
distance $d$ as compared to the separation of the stars in a binary system.}
lens solutions that are degenerate solutions when looking merely to the observed light curves. 

In the present paper, also in view of the fact that highly magnified events 
considered in our previous works are not so frequent,
we extend the analysis of the expected polarization signal for single lens events 
towards the Galactic bulge, covering the full range of microlensing parameters.

Concerning the polarization mechanism in operation, it is clear from our analysis
\citep{Ingrosso12} that, due to paucity of hot stars in the Galactic bulge, 
the source stars with the highest chance to give rise to a non negligible
polarization signal are cool, evolved giants, for which the photon scattering on dust grains
is the leading polarization mechanism. 

Therefore, following the model by \cite{Simmons02}, briefly summarized in Sect. 2, 
we assume that dust grains may form in the stellar atmospheres, beyond the 
distance $R_h$ greater than the source radius $R_S$, at which the gas temperature 
becomes lower than the dust sublimation temperature. 
{
The model is characterized by the presence of an internal cavity 
(between $R_S$ and $R_h$) devoid of dust, giving rise during a microlensing event
to two different polarization profiles, as a function of the time $t$, for {\it bypass}  and 
{\it transit} events \citep{Simmons95a,Simmons95b}. 
Indeed, when the trajectory (projected in the lens plane) of the cavity remains 
outside the lens ({\it bypass} events), one obtains a bell-like polarization profile with 
the peak occurring at the instant of maximum magnification $t_0$. 
When a part of the cavity is aligned with the lens and the observer ({\it transit} events),
the polarization curve presents two maxima and one minimum (in correspondence to $t_0$). 
The polarization signal gets the two maximum values when the cavity enters and exits 
the lens.}

Note that a major advantage of considering the polarization (namely the microlensing of the 
envelope instead of the stellar disk) is that the probability for observing finite 
source effects in microlensing events increases \citep{Yoshida}.  
Indeed, in the case of {\it bypass} events, for which the
trajectory of the source disk 
does not intersects the lens, 
the finite source effects, negligible on the event light-curves
- that are not sensitive to the limb-darkening phenomena
\citep{GG99,HS00,Abe2003} -  
produce a polarization signal that constitute, therefore, a
unique probe of the stellar atmosphere.
In {\it transit} events for which the source disk transverses the lens, 
the finite source effects manifest themselves in the polarization variability
when the internal cavity transits the lens, while in the light-curves their effects 
may appear or not depending on 
the sensitivity of the light-curve to the limb-darkening phenomenon
\citep{GG99,HS00,Abe2003}.

{
Our present analysis of the polarization runs along two parallel paths, 
the first being a study of the signal present in microlensing events
generated by using a Monte Carlo code. This preliminary study allows 
us to clarify the dependence of the polarization signal on the 
microlensing parameters and the physical parameters of the source stars.

Then we consider the set of OGLE-III events relative to the 2001-2009 
observational campaign towards the Galactic bulge recently reported in 
\cite{Wyrzykowski2014}
with an evaluation of the expected polarization signal. This analysis is however 
complemented by a Monte Carlo study, 
since the observational data alone do not allow us to
completely characterize the polarization signal.

The overall analysis eventually leads us to determine the physics that can be extracted by
the study of the polarization as well as the optimal
observational strategy for forthcoming microlensing events.
Indeed, our} final aim is to design an observational strategy which allows us to maximize the 
chance of positive polarization measurements with an observational programme based on 
the current microlensing surveys. To this aim we consider the OGLE-III
event catalog recently presented by \cite{Wyrzykowski2014}.
All these available data allow to compute the maximum
polarization degree in each of the observed events and the best time to 
perform a polarization measurement in an ongoing event, 
once an alert system has predicted the instant of the occurrence 
of the magnification peak.
This, in turn, also makes possible to estimate that for a few events
per year (about 1 percent of the OGLE-III events) the FORS2 polarimeter 
on VLT telescope, in 1 hour of 
integration time, may allow to obtain a positive polarization measurement with 
$P_{\rm max} > 0.1$ percent (see Sect. 5).

\section{Polarization model}

Following the approach in \cite{Chandra60}, we consider the linear polarization of the
star light scattered in a stellar atmosphere. We define the intensities $I_l(\mu)$ and 
$I_r(\mu)$ emitted in the direction making an angle $\chi=\arccos(\mu)$ with the normal
to the star surface and polarized as follows: $I_l(\mu)$ is the intensity in the plane 
containing the line of sight and the normal, $I_r(\mu)$ is 
the intensity in the direction perpendicular to this plane 
(light propagates in the direction ${\bf r \times l}$). 

To calculate the polarization of a star with center at projected position $(p_S,\varphi_S)$ 
in the lens plane we integrate the unnormalized Stokes parameters and the flux over the star 
disk
\citep{Simmons95a,Simmons95b,Agol96}
\begin{eqnarray}
F   = F_0 \int_{0}^{2\pi} \int_0^{\infty} A(p,\varphi) ~ I_+(\mu)               ~ p dp ~ d \varphi~,
\label{flux} 
\end{eqnarray}
\begin{eqnarray}
F_Q = F_0 \int_{0}^{2\pi} \int_0^{\infty} A(p,\varphi) ~ I_-(\mu)~\cos 2 \varphi ~p dp ~ d \varphi~,
\label{fq} 
\end{eqnarray}
\begin{eqnarray}
F_U = F_0 \int_{0}^{2\pi} \int_0^{\infty} A(p,\varphi) ~ I_-(\mu)~\sin 2 \varphi ~p dp ~ d \varphi~,
\label{fu} 
\end{eqnarray}
where
$F_0$ is the unamplified star flux, $A(p, \varphi)$ the point source magnification and
\begin{eqnarray}
I_+(\mu) = I_r(\mu)+I_l(\mu)~, 
\end{eqnarray}
\begin{eqnarray}
I_-(\mu)=I_r(\mu)-I_l(\mu) ~.
\end{eqnarray}

{ As usual, the polarization degree is given by $P = (F_Q^2+F_U^2)^{1/2}/{F}$.}

We use a coordinate system with the lens at the origin $O$ and with the
source trajectory along the $x$ axis. The location of a point $(p,\varphi)$
on the star surface is determined by the 
distance $p$ from the star center and by the angle $\varphi$ formed with the $Ox$ axis. 
Since we are considering single-lens events,
for each surface element on the stellar 
disc the magnification is given by 
\citep{Einstein36,Pacz86} 
\begin{equation}
A = \frac{u^2+2}{u \sqrt{u^2+4}}~,
\label{A-1lente}
\end{equation}
where $u$ is the distance, in units of the Einstein radius $R_E$, 
between the considered surface element and the lens position 
\begin{equation}
u = \sqrt{p_S^2 + p^2 - 2 p p_S \cos (\varphi-\varphi_S)} \ .
\end{equation}
{Here, $R_E=[(4Gm_L/c^2)(D_L/D_S)(1-D_L/D_S)]^{1/2}$,
being $m_L$ the lens mass and $D_S$ ($D_L$) the source (lens) distance from the observer.}  

Clearly, by taking into account the finite source effect,
the overall source magnification is obtained by integration on the distance 
${\bar R}_S=(R_S/R_E)(D_L/D_S)$ 
{ (the physical radius $R_S$, in units of $R_E$, projected in the lens plane)}
\begin{equation}
\langle A \rangle = \frac
{ \int_{0}^{2\pi} \int_0^{{\bar R}_S} A(p,\varphi) ~p dp ~ d \varphi}
{ \int_{0}^{2\pi} \int_0^{{\bar R}_S}             ~p dp ~ d \varphi}~.
\end{equation}

The light-curve and polarization profiles as a function of the time $t$ are obtained  
by specifying in the above equations the time-dependent
position of the source star center 
\begin{equation}
p_S = \sqrt{u_0^2 + [(t-t_0)/t_E]^2} \ \ , \ \ \ \varphi_S = \arctan \frac{u_0}{(t-t_0)/t_E} \ ,
\end{equation}
{ where $u_0$ is the lens impact parameter, $t_0$ the maximum magnification time and $t_E$
the Einstein time.}  

{ 
The explicit form of the intensities $I_+(\mu)$ and $I_-(\mu)$ is given in
\cite{Simmons02} and \cite{Ignace06}. 
It turns out that the polarization $P$ linearly depends on the dust optical depth 
\begin{equation}
\tau = n_h \sigma R_h/(\beta-1) \ ,
\label{tauwind}
\end{equation} 
in the limit of $\tau <<1$. Here,
$\sigma$ is the scattering cross-section (evaluated in the dipole approximation)
of photons off dust grains, 
$R_h$ is the distance at which dust grains may form in the stellar wind
and the scatterers are taken to have a number density decreasing with the 
distance (from the $n_h$ value) 
with a power law of esponent $\beta$.
}


We estimate the distance $R_h$ according to simple energy balance 
criteria by considering the balance between the energy absorbed and emitted
 by a typical dust grain as a function of the distance 
$r$ from the star center
\begin{equation}
\int_0^\infty 
F^S_\lambda(r) \pi a^2 Q_\lambda d\lambda = \int_0^\infty 4 \pi a^2 \pi  
B_\lambda[T(r)] Q_\lambda d\lambda~,
\label{bilancio}
\end{equation}
where $F^S_{\lambda} (r)$ is the stellar flux at distance $r$
\begin{equation}
F^S_\lambda(r) = \left(\frac{R_S}{r}\right)^2 \pi B_\lambda(T_{\rm eff}) ~.
\label{flux_p}
\end{equation}
Here $T_{\rm eff}$ is the effective temperature of the source star,
$B_\lambda$ the black body emissivity at the wavelenght $\lambda$, 
$Q_\lambda$ the grain absorption efficiency,
$T(r)$ the dust temperature at distance $r$ and $a$ the dust grain size. 
This calculation assumes that the heating by non radiative processes
and by the diffuse radiation field is negligible so that we limited ourselves to compute 
$Q_\lambda$ for a typical particle size distribution \citep{Mathis}
with optical constants derived by \cite{Draine}.
Specifically, the numerical values for $R_h$ are obtained by using 
eq. (\ref{bilancio}) with $T_h \equiv T(R_h) \simeq 1400 \ ^0$K.

{Concerning the dust optical depth, it was found that}  
the dependence of $\tau$ from stellar and wind parameters 
can be approximated by \citep{Ignace08}  
\begin{eqnarray}
\tau  = 2 \times 10^{-3} \eta {\mathcal K} 
\left(\frac{\dot{M}} { 10^{-9}~M_{\odot}/{\rm yr}}\right)
\left(\frac {30~{\rm km/s}} {v_{\infty}}\right)
\left(\frac {24R_{\odot}} {R_h} \right)~,
\label{tau} 
\end{eqnarray}
where $\eta \simeq 0.01$ is the dust-to-gas mass density ratio, 
${\mathcal K} \simeq 200$ cm$^2$ g$^{-1}$ is the dust opacity at
$\lambda > 5500$ \AA, $\dot{M}$ is the mass-loss rate and 
$v_{\infty}$ is the asymptotic wind velocity. Observational data indicate 
that $v_{\infty}$ is related to the
escape velocity $v_{\rm esc}=\sqrt{2GM_S/R_S}$, depending on the source spectral and luminosity 
class. In particular, $v_{\infty} \simeq 0.2 v_{\rm esc}$ 
in the case of AGB stars 
\citep{Marshall} that are the sources giving the highest polarization signals.
 
We can relate $\dot{M}$ in eq.(\ref{tau}) to the 
luminosity $L$, gravity $g$ and radius $R$ of the magnified source star. 
Indeed, it is well known that from main sequence to AGB phase stars, 
the mass-loss rate $\dot{M}$ increases by several orders of magnitude.
Here, we adopt the simple empirical mass-loss rate relation \citep{Reimers75}
\footnote{Here we note that several different mass-loss rate relations are present 
in the literature obtained  by fitting
observational data - see, e.g., \cite{Catelan00} - and by numerical models \citep{Ferguson05}.}
\begin{equation}
\dot{M} = \eta_{R}  \ 4 \times 10^{-13} \ \frac{ (L/L_{\odot})}
{ (g/g_{\odot}) \  (R/R_{\odot}) } ~ (M_{\odot}/{\rm yr})~.
\label{dotM}
\end{equation}
where $L_{\odot}$, $g_{\odot}$ and $R_{\odot}$ are luminosity, gravity and radius for the Sun,
respectively, and $\eta_{R}$ is a free parameter that is estimated by a best fit procedure to 
the observational data, being $\eta_R \simeq 0.3$ for late-type main sequence stars
and $\eta_R \simeq 3$ for red giant stars. 

By assuming the relation in eq.(\ref{dotM}), values of $\dot{M}$ 
in the range $(10^{-12} - 10^{-8})~M_{\odot}~{\rm yr}^{-1}$ 
are obtained for typical stars evolving from main sequence 
to red giant star phases. 

{
Before closing this section we remark that, as it is well known,
in analyzing microlensing observations, the parameters $t_0$, $u_0$, 
$t_E$ (and, possibly, ${\bar R}_S$ for events with large finite source effects) 
are determined by fitting each light-curve with the \cite{Pacz86} law, 
while the distances $D_S$, $D_L$ and $R_E$ remain undetermined.
Then, also given the $R_S$ and $R_h$ physical radii,  
the adimensional distances ${\bar R}_h = (R_h/R_E) (D_L/D_S)$ 
and ${\bar R}_S$ (for events with negligible finite source effects)
remain unkwown, making not possible to compute the polarization signal 
raised in the microlensing event. 
The way out we adopt in the analysis of the observd OGLE-III events (Sect. 4.2)
is to use a Monte Carlo code to determine the more likely
values of $D_S$, $D_L$ and $R_E$ corresponding to the event $t_E$ value.}

\section{Polarization results}

\begin{figure}
\includegraphics[width=80mm]{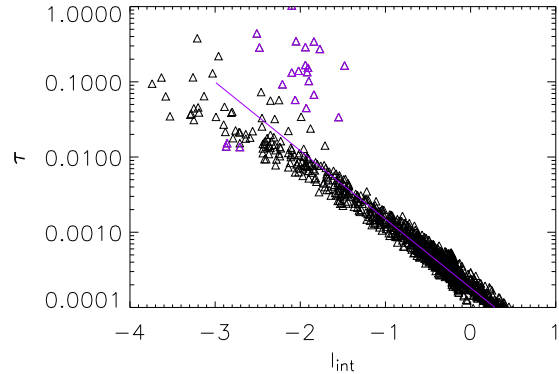}
\caption{The ratio $P(t_0)/\tau$ in percent at the time $t_0$ of maximum magnification 
is given (solid lines) as a function ${\bar R}_h/u_0$, assuming 
${\bar R}_h/{\bar R}_S = 2, \ 5$ (from the bottom).
For {\it bypass} events (${\bar R}_h < 0.75 u_0$), $P(t_0)/\tau$ is the maximum of the
polarization variability as a function of the time $t$. 
{In the case of {\it transit} events (${\bar R}_h > 0.75 u_0$), $P(t_0)/\tau$ is a local minimum of the 
polarization curve and the maximum value $P(t_h)/\tau \simeq 12.5$  occurs twice at the $t_h$ instants given by eq. 
(\ref{transit_time}).}}
\label{fig1}
\end{figure}

{

In this section we describe the dependence of the intensity and shape of the 
polarization signal from the microlensing parameters 
$u_0$ and $t_E$, and from the parameter ratios ${\bar R}_h/u_0$ and 
${\bar R}_h/{\bar R}_S$.

At first we focus on the polarization signal $P(t_0)/\tau$ at the instant $t_0$ of 
maximum magnification, which is independent from the value of $t_E$. 
Fig. \ref{fig1} shows 
\footnote
{ We point out that in our previous analysis 
\citep{Ingrosso12} there was a mistake in using eq. (13) in \cite{Simmons02},
that gives rise to a different $P(t_0)/\tau$ profile (shown in Fig. \ref{fig1})
with respect to that published in Fig. 3 of \cite{Ingrosso12}.}
the ratio  $P(t_0)/\tau$ in percent as a function of ${\bar R}_h/u_0$.} 
The curves are given for $\beta=2$ and for two different values of 
${\bar R}_h/{\bar R}_S=2$ and 5. We have verified that
in the limit $u_0<<1$, the $P(t_0)/\tau$ profiles weakly depend on 
${\bar R}_h/{\bar R}_S$ and any curve is inside the region bounded by the 
two curves plotted  in Fig. \ref{fig1}.
Therefore, the polarization level 
at the time $t_0$ depends only on the ratio ${\bar R}_h/u_0$, 
assuming the maximum value of about $P(t_0)/\tau \simeq 12.5$ percent 
when ${\bar R}_h / u_0 \simeq 0.75$. 

The effect of varying the model parameter $\beta$ in eq. (\ref{tauwind}) is shown in 
Fig. 3 in \cite{Ingrosso14}, where one can see that increasing $\beta=2, \ 3, \ 4$,  
the $P(t_0)/\tau$ values increase from 12.5 percent up to about 25 percent. 
However, since $P(t_0)/\tau$ rapidly decreases 
outside the interval $0.1 < {\bar R}_h/u_0 < 2.2$,
it is clear that the presence of a large enough polarization signal at $t_0$ 
requires that microlensing and stellar parameters $u_0$ and ${\bar R}_h$ be fine tuned.


The ratio ${\bar R}_h/u_0$, not only determines the value of the polarization 
signal at $t_0$, but also the shape of the polarization profile $P(t)/\tau$ as a function of 
time, and in particular the possible existence of two polarization peaks at 
symmetrical positions with respect to $t_0$.

Indeed, when ${\bar R}_h/u_0 \ut < 0.75$, i.e. for {\it bypass} events
\begin{figure}
\includegraphics[width=90mm]{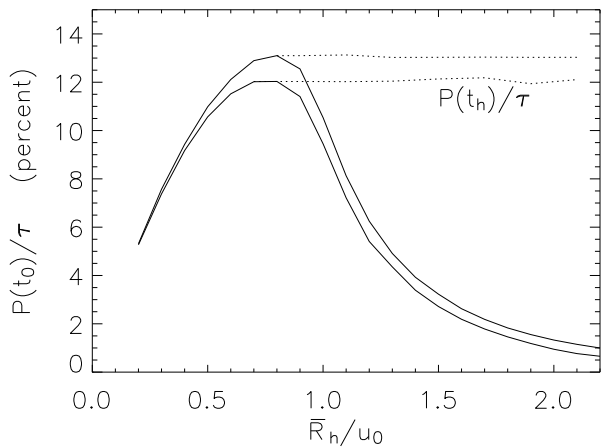}
\caption{Assuming constant parameter values $u_0=0.09$, $t_E=60$ day and 
${\bar R}_h/{\bar R}_S = 5$, the $P(t)/\tau$ polarization curves are shown as a function of 
$(t-t_0)/t_E$, 
for increasing values of ${\bar R}_h/u_0 = 0.35, \ 0.75, \ 1.5,  \ 2.5$, corresponding to
curves labelled a), b) c) and d), respectively.
Full lines - a) and b) - are  {\it bypass} events, dotted lines - c) and d) -
are {\it transit} events.}
\label{fig2}
\end{figure}
- for which the trajectory of the internal star cavity (devoid of dust grains) remains 
outside the lens - one obtains a bell-like  polarization profile with the peak occurring at $t_0$.

For ${\bar R}_h/u_0 \ut > 0.75$, i.e. for {\it transit} events - where a part of the internal cavity is
aligned with the lens and the observer - the polarization curve present two maxima and one minimum
(in correspondence to $t_0$). 
The polarization signal gets the two maximum 
values when the internal cavity enters and exits the lens.

The different behaviour of the polarization profiles for {\it bypass } and 
{\it transit} events is clarified in Fig. \ref{fig2} where, by taking constant values
for the parameters $u_0$ and $t_E$, we show the $P(t)/\tau$ profiles 
as a function of the ''adimensional time'' $(t-t_0)/t_E$, for increasing values of ${\bar R}_h/u_0$.
As one can see in Fig. \ref{fig2}, the same maximum value of $P(t)/\tau \simeq 12.5$ percent
is achieved for the simulated event labelled by b) 
(the limiting {\it bypass}/{\it transit} event with ${\bar R}_h/u_0 = 0.75$)
and for events c) and d), that are two {\it transit} events with 
${\bar R}_h/u_0 = 1.38$ and $2.06$, respectively. 
{The effect of varying in Fig. \ref{fig2} the parameter $t_E$  
is to change the time scale of the polarization variability, while the event nature
({\it transit} or {\it bypass}) remains unchanged.}

This behaviour of the polarization profiles is indeed expected 
since for {\it transit} events the central value $P(t_0)/\tau$ 
corresponds to the minimum of the polarization curve - 
with intensity depending on ${\bar R}_h/u_0$  as shown in Fig. \ref{fig1} - 
and, for increasing $t$, there always exists along the star trajectory 
two positions of the source star center at which ${\bar R}_h/u_h \simeq 0.75$,
beeing $u_h=\sqrt{u_0^2+[(t_h-t_0)/t_E]^2}$.  
This clearly implies that the two polarization peaks occour at times 
\begin{equation}
t_h =   t_0 \pm t_E  \sqrt{ \left( \frac{{\bar R}_h}{0.75} \right)^2-u_0^2}~.
\label{transit_time}
\end{equation}

{It goes without saying that the apparent source star luminosity at $t_h$ decreases  
with respect to the value at the maximum magnification time $t_0$, therefore
making more hard performing a measurement of the polarization signal 
in events with relatively large values of ${\bar R}_h/u_0$.}

\section{Observational outlook}

This section aims to present a discussion of the observational perspectives to 
effectively perform a polarization measurement during an ongoing microlensing event.
 
Very recently, the OGLE collaboration presented the largest and more comprehensive
catalog of microlensing events ever constructed \citep{Wyrzykowski2014}. 
The sample of single lens events comprises 3718 events towards the Galactic bulge for  
years 2001-2009 (OGLE-III campaigns), 
with 1410 events not detected before in real time by the Early Warning 
System of the OGLE experiment. The observed light-curves have been re-analyzed and new 
values of the microlensing parameters have been obtained. The microlensing event detection 
efficiency was also determined as a function of the Einstein time.



{
We first make a study (in Sect 4.1)  
of the maximum polarization signal occurring  
in microlensing events generated by means of a Monte Carlo code.
Then, we estimate (in Sect 4.2) the polarization degree that would occur
in the real OGLE-III events. 

We remark that the Monte Carlo and OGLE-III analyses complement each other.
In fact, starting from the full sample of Monte Carlo generated events,
we make use of the OGLE-III event detection efficiency,
so to draw our final sample of ''potentially observable'' events.
In turn, the Monte Carlo study allows us to determine in the OGLE-III analysis
the unkwon distance scales $D_S$, $D_L$ and $R_E$, making possible to compute
the polarization signal raised in the event.}


\subsection{Monte Carlo analysis}

We generated a sample of microlensing events with source stars in the Galactic bulge and 
lenses either in the bulge or in the disk of the Galaxy. For the sake of semplicity, we 
assume sources at distance $D_S=8$ kpc and the line-of-sight to the Galactic center ($l=0$, 
$b=0$). The mass and the distance of the lenses and the velocity of both sources and lenses
are selected, using a model for the distribution of the stars in the Galaxy, 
by means of the Monte Carlo technique \citep{Ingrosso06,Ingrosso07,Ingrosso09}.  


As a first step, we have verified that the obtained distribution of the Einstein time $t_E$ is 
in agreement with the observational results, as given by Fig. 7 in \cite{Wyrzykowski2014}. 
Then, we use the OGLE-III event detection efficiency as a function of $t_E$ 
(given in Fig. 8 of the above cited paper) to select the ``potentially observable'' events,
starting from the full sample of generated events.

{
In the Monte Carlo analysis we draw the source stars according
to a synthetic luminosity function generated
using the IAC-Star CMD code \footnote{http://iac-star.iac.es/cmd/www/form.htm}.
Here we assume that the Galactic bulge stars were generated by a
single star-burst that occurred about 13 Gyr ago in a 
solar metallicity gas cloud and  adopting a \cite{Kroupa} initial mass function.
Then, for each star of our adopted catalogue we compute the cavity radius $R_h$ 
by using eqs. (\ref{bilancio}) and (\ref{flux_p}) with the appropriate values
of $R_S$ and $T_{eff}$.  
\footnote{ 
Based on the model we use for $R_h$ and $T_h$, 
we find the scatter plot $R_h/R_S$ vs $T_{\rm eff}$ to follow the analytical relation
${R_h}/ {R_S} = 0.45 ({T_{\rm eff}}/{T_h})^{2.5}$ 
found by \cite{lamcas}.} 
}

We explore the full range of stellar luminosities and temperatures
allowing  the color index $(V-I)_{\rm int}$ to vary uniformly 
in the range $-5 < I_{\rm int} < 0.5$, also including source stars 
that are not present in the OGLE-III catalog of observed sources.
{ 
Here and in the following we indicate with the subscript ''$_{\rm int}$'' dereddened 
quantities of the unlensed source star.
\footnote{
We find a linear relation between the intrinsic color index $(V-I)_{\rm int}$ 
and the effective star temperature $T_{\rm eff}$, which is consistent with the
widely accepted analytical approximations in the literature 
\citep{sekiguchi}.}
}


Given the physical parameters of the source stars, 
by using eqs. (\ref{tau}) and (\ref{dotM})
we estimate the dust optical depth $\tau$ and in turn the maximum polarization degree 
given by $P_{\rm max} \equiv P(t_{\rm max}) \simeq 12.5 ~ \tau$ percent,
being $t_{\rm max} = t_0$ for {\it bypass} events and 
$t_{\rm max} =  t_h$ for {\it transit} events (see Sect. 3.).

\begin{figure}
\vspace{5.5cm} \includegraphics{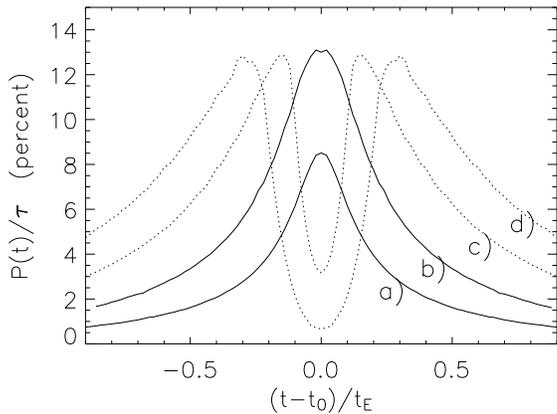}
\caption{Scatter plot of the dust optical depth $\tau$ vs the source star
intrinsic magnitude $I_{\rm int}$.}
\label{fig3}
\end{figure}

The corresponding values of $\tau$ are shown 
in Fig. \ref{fig3} where we give a scatter plot of $\tau$ vs $I_{\rm int}$. 
In the same figure, for comparison purposes, we show a linear
fit to the simulated data and, as colored triangles, the points with
$\tau$ values above two standard deviations with respect to the linear fit.


\begin{figure}
\vspace{5.5cm} \includegraphics{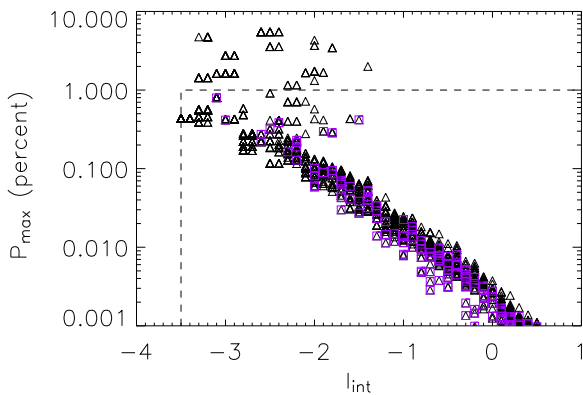}
\caption{Scatter plot of $P_{\rm max}$ in percent vs the unmagnified 
star magnitude $I_{\rm int}$ for simulated {\it transit} (triangles) and 
{\it bypass} (purple squares) events. The dashed lines delimit 
the region containing the results for the OGLE-III events (see Sect. 
\ref{sect_OGLE_events}).}
\label{fig4}
\end{figure}

\begin{figure}
\vspace{5.5cm} \includegraphics{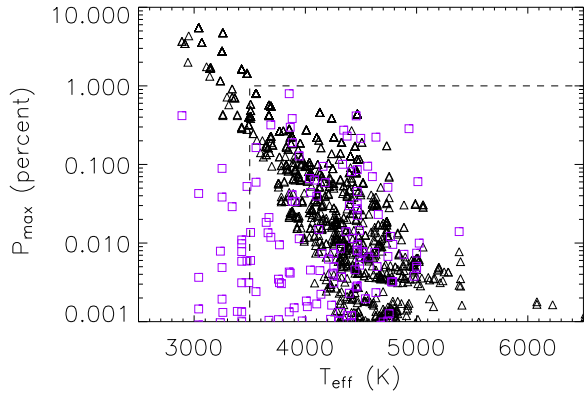}
\caption{Scatter plot of $P_{\rm max}$ in percent vs the source temperature $T_{eff}$ 
for {\it transit} (triangles) and 
{\it bypass} (purple squares) events. The dashed lines delimit 
the region containing the results for the OGLE-III events (see Sect. 
\ref{sect_OGLE_events}).}
\label{fig5}
\end{figure}

\begin{figure}
\vspace{5.5cm} \includegraphics{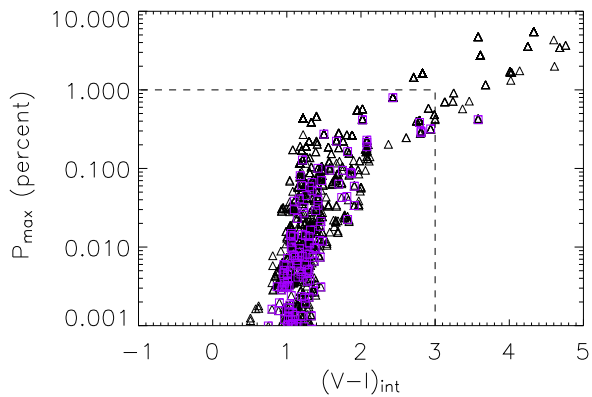}
\caption{Scatter plot of $P_{\rm max}$ in percent vs the color index
$(V-I)_{\rm int}$ for simulated {\it transit} (triangles) and 
{\it bypass} (purple squares) events. The dashed lines delimit 
the region containing the results for the OGLE-III events (see Sect. 
\ref{sect_OGLE_events}).}
\label{fig6}
\end{figure}

{
In Figs. \ref{fig4}, \ref{fig5} and \ref{fig6}
we give $P_{\rm max}$ as a function of $I_{\rm int}$, 
$T_{\rm eff}$ and  $(V-I)_{\rm int}$ of the source stars, respectively. 
As one can see, we have $P_{\rm max} < 1$ percent for red giants with 
$(V-I)_{\rm int } < 3$, which corresponds to 
$I_{\rm int}>-3.5$ and  $T_{\rm eff}>  3500$ K 
(events inside the regions delimited by dashed lines).
There are, however, a few events with $1 < P_{\rm max} < 10$ percent,
characterized by $(V-I)_{\rm int }>3$ and $T_{\rm eff}< 3500$ K, 
corresponding to source stars in the AGB phase.
These stars, that are rather rare in the Galactic bulge, 
have not been as yet sources of microlensing events observed 
in the OGLE-III campaign.

In this respect,
the expected significant increase in event rate 
by the forthcoming new generation Bulge microlensing surveys
(both ground-based, as KMTNet \citep{KMT14},
and from space, EUCLID \citep{Penny} and WFIRST 
\citep{Yee}
open the possibility to develop an alert system
able to trigger polarization measurements
in ongoing microlensing events with very bright sources.}

\subsection{OGLE-III events}\label{sect_OGLE_events}

We now consider the sample of single lens OGLE-III events 
recently re-analyzed by \citep{Wyrzykowski2014}.
Our aim is to compute the expected polarization signal in 
2614 events (out of the 3718 total number)
for which line-of-sight, microlensing parameters and 
baseline magnitudes in V and I bands of the source stars are given.

{
In the present analysis we assume that the baseline
magnitude and color are reliable estimate
of the source magnitude and color, namely
we are assuming blending is negligible.
Despite this assumption is in general questionable, it is
however reasonable at least for the brighest stars in the sample,
that our analysis favours as the best candidates to produce appreciable 
polarization signals.}

{
We derive the physical parameters of each microlensed 
source by assuming that the interstellar extinction $A_I$ and the 
reddening $E(V-I)$ of each source star are those measured by \cite{Nataf}
for the OGLE field of view towards the Galactic bulge nearest to the event line-of-sight.
In this respect, we point out that attributing the average
values of $A_I$ and $E(V-I)$ in each field of view to the single sources may 
alterate the true values of the stellar physical parameters, thereby producing 
a difference in the estimated  values of $\tau$ and  ultimately of $P_{\rm max}$. 
Indeed, this is exactly what happened in the case of the event
OGLE-2011-BLG-1101/MOA-2011-BLG-325 \citep{Ingrosso12} for which 
$T_{\rm eff} \simeq 5100$K if it is estimated by using the catalog in
\cite{Nataf} while $T_{\rm eff} \simeq 3800$ in \cite{Choi}.  
}

\begin{figure}
\vspace{5.5cm} \includegraphics{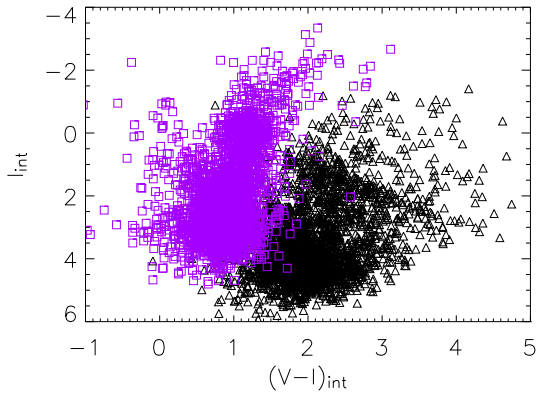}
\caption{Scatter plot of $I_{\rm int}$ vs $(V-I)_{\rm int}$ (purple squares) for the 
stars in the OGLE-III catalog \citep{Wyrzykowski2014}
corrected for absorption and reddening.
For comparison, we give also the scatter plot of the observed quantities  
$I$ and $(V-I)$ (triangles).} 
\label{fig7}
\end{figure}

In Fig. \ref{fig7} we give the scatter plot of the 
intrinsic magnitude
$I_{\rm int} =I-A_I$ vs the color index $(V-I)_{\rm int} = (V-I)+E(V-I)$.
We observe that the distribution of the magnitude for the microlensed source stars 
is different from that of the bulge stars as an effect of the 
event selection criteria \citep{Wyrzykowski2014}. 

The effective temperature $T_{\rm eff}$ of each source star is then evaluated since is kwown
the relation between intrinsic color index $(V-I)_{\rm int}$ and $T_{\rm eff}$.
In turn, the stellar radius is calculated through the relation  
$ L_{\rm Bol} = 4 \pi \sigma R_S^2 T^4_{\ \rm eff}$, and finally 
the microlensed star mass is selected by searching in the IAC catalog 
a star with the assigned values of $I_{\rm int}$, $T_{\rm eff}$ and $R_S$.
This allows to estimate the mass-loss rate through eq. (\ref{dotM}) and
finally $\tau$ as given by eq.(\ref{tau}).

{
The OGLE-III event paramaters by themselves do not allow us a complete characterization
of the expected polarization signal, for which we need, in particular, a
full knowledge of the microlensing event parameters. 
For this purpose we here make use once
again of the Monte Carlo model constrained by the observational data. }
{ Assuming a model for the distribution of sources and lens in the Galaxy,
we generate, a synthetic microlensing event catalog, by selecting the source 
(lens) distance $D_S$ ($D_L$), the lens mass $m_L$ and the relative (projected 
in the lens plane) velocity $v_{\perp}$ between the source and the lens  
\citep{Ingrosso06,Ingrosso07,Ingrosso09}.
Then, for each OGLE-III event, we select within the catalog, the 
event having the ratio between $R_E$ and $v_{\perp}$ (being $t_E = R_E / v_{\perp}$)
nearest to the observed one.
In this way we are able to evaluate the adimensional ${\bar R}_S$ and ${\bar R}_h$ radii.}

For each of the considered OGLE-III events
we thereby 
compute the maximum polarization degree $P_{\rm max} \equiv P(t_{\rm max})$:
$t_{\rm max}=t_0$ for {\it bypass} events, $t_{\rm max}=t_h$
                  for {\it transit} events.
All events have $P_{\rm max}<1$ percent and only a few dozens
have $0.1 < P_{\rm max} < 1$ percent. The obtained scatter plots 
of $P_{\rm max}$ vs $I_{\rm int}$, $T_{\rm eff}$ and
$(V-I)_{\rm int}$ (not given) are similar to those obtained for Monte Carlo 
events shown in the Figs. \ref{fig4}, \ref{fig5} and \ref{fig6}, 
with data points inside the region $P_{\rm max} <1$ percent, 
$I_{\rm int} > -3.5$, $T_{\rm eff} > 3500$K and $(V-I)_{\rm int}<3$, respectively.
The lack of events in the OGLE-III catalog with $P_{\rm max}>1$ percent
is clearly due to the paucity of bulge AGB stars implying that none of them
has been microlensed during the 2001-2009 campaigns.

\begin{figure}
\vspace{5.5cm} \includegraphics{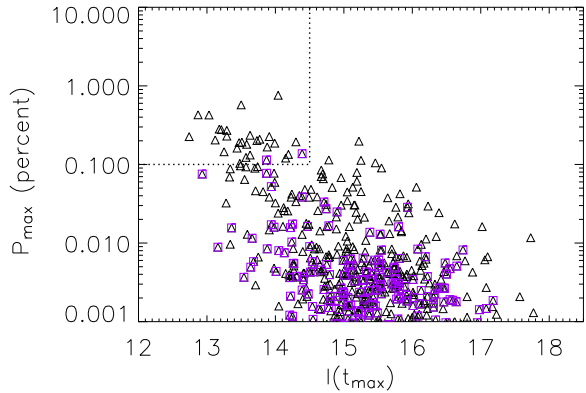}
\caption{Maximum polarization expected in OGLE-III events. 
$P_{\rm max}$ in percent is given vs the apparent magnitude $I(t_{\rm max})$.
either for {\it transit} (triangles) and {\it bypass} (purple squares) 
events.
The region enclosed in dashed lines shows the more favourable events 
(about 1 percent of the OGLE-III events) for polarization measurements, 
assuming 1 hour integration time with FORS2 at ESO-VLT telescope.}
\label{fig8}
\end{figure}

In Fig. \ref{fig8} the maximum polarization degree $P_{\rm max}$ 
expected for the OGLE-III events is given as a function of the stellar apparent magnitude
$I_{\rm max}$ at the time instant of maximum polarization.
The region enclosed in dotted lines shows the more favourable events 
(about 1 percent of the OGLE-III events) for polarization measurements, 
assuming 1 hour integration time with FORS2 at ESO-VLT telescope
\citep{Ingrosso12}.

\begin{figure}
\vspace{5.5cm} \includegraphics{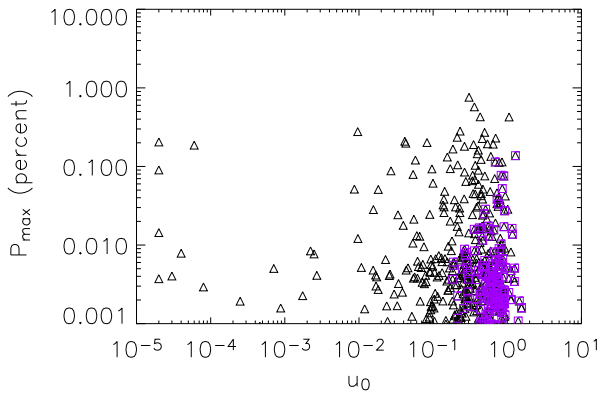}
\caption{Scatter plot of $P_{\rm max}$ vs the impact parameter $u_0$
either for {\it transit} (triangles) and {\it bypass} (purple squares) 
OGLE-III.}
\label{fig9}
\end{figure}

\begin{figure}
\includegraphics[width=90mm]{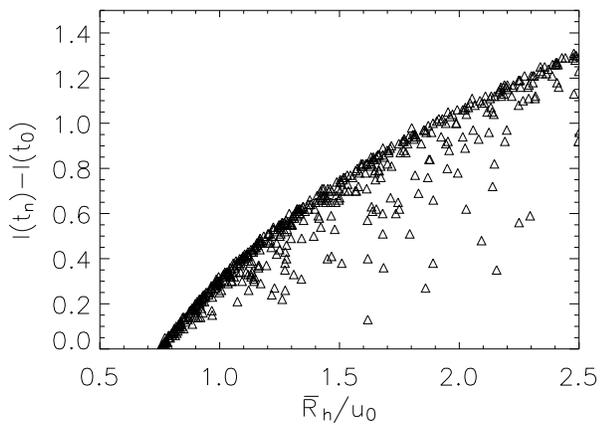}
\caption{For {\it transit} events, the increase of source magnitude 
$I_{\rm int}(t_h)-I_{\rm int}(t_0)$  at the instant $t_h$ of maximum polarization 
with respect the instant $t_0$ of maximum magnification is shown as a function of ${\bar R}_h/u_0$.}
\label{fig10}
\end{figure}

The event distribution with the impact parameter is shown in Fig. \ref{fig9},
where one can see that {\it transit} events have higher polarization signals with respect to
{\it bypass} ones.
This figure also shows that the same polarization level  
is achieved in events with low and high magnification.
Finally, in Fig. \ref{fig10}, for {\it transit} events, we show the increase
of apparent magnitude $\Delta I = I(t_h)-I(t_0)$
between the time instant $t_h$ of maximum polarization and $t_0$
of maximum magnification: $\Delta I$ increases with increasing 
${\bar R}_h/u_0$ and this makes more hard to measure the polarization 
in events with large values of ${\bar R}_h/u_0$.

\section{Summary and conclusions}

The study of the OGLE-III microlensing events towards the Galactic bulge
recently reported by \cite{Wyrzykowski2014}, integrated by a Monte Carlo
analysis, allows us to compute the expected polarization signal 
for a large number of events (2616 out of 3718 events) for which all the 
physical parameters of the source stars can be determined.

We focus on cool giant source stars, as they provide a higher level of polarization,
and we follow the model by \cite{Simmons02} for which the polarization degree $P$ 
is proportional to the dust optical depth $\tau$. This quantity, computed by using
the relations in eqs. (\ref{tau}) and (\ref{dotM}), is found  
in the range $10^{-4} < \tau < 10^{-1}$ for typical stars evolving from (late type)
main sequence  to red giant phase. An important model parameter is the distance $R_h$
in the stellar envelope at which dust grains may form. It depends on radius 
and effective temperature of the source star as shown in Sect. 4.1. 

With this in hand, we can evaluate the polarization profile $P(t)$ during 
a microlensing event. It turns out that the ratio ${\bar R}_h/u_0$ determines 
both the strength (apart the dependence of $P$ on $\tau$) and the 
shape of the expected polarization signal.

In particular, when ${\bar R}_h/u_0 < 0.75$, i.e. for {\it bypass} events,
the polarization $P(t)$ shows a bell-like profile with the peak occurring at $t_0$. 
As shown in Fig. \ref{fig1}, the actual value of $P_{\rm max}(t_0)/\tau$ 
depends on the ratio ${\bar R}_h/u_0$, being  
$P_{\rm max}(t_0)/\tau \simeq 12.5$ percent 
at ${\bar R}_h/u_0 \simeq 0.75$. 

For ${\bar R}_h/u_0 >  0.75$, i.e. for {\it transit} events, 
the polarization curve presents a central minimum (in correspondance to $t_0$),
while two maxima occur at the time instants 
$t_h =  t_0 \pm t_E  \sqrt{({\bar R}_h/0.75)^2-u_0^2}$ at which 
the  internal cavity (of the stellar atmosphere devoid of dust, projected in the lens plane)
enters and exits the lens. The maximum polarization degree is again
$P_{\rm max}(t_h)/\tau \simeq 12.5$ percent, corresponding to a lens-source projected
distance $u_h = 0.75 u_0$.

Clearly, with this in mind, one can predict in advance for which events 
and at which time the observing resources may be focused to make a polarization
measurement in an ongoing microlensing event.

The Monte Carlo analysis 
shows how $P_{\rm max}$ depends on the stellar parameters 
(see Figs. \ref{fig4}, \ref{fig5} and \ref{fig6}): 
$P_{\rm max}$ increases from (late type) main sequence to AGB star phase
increasing from $10^{-3}$ up to values of a few percent.
Our analysis also shows that the {\it transit} events have a higher polarization degree 
with respect to {\it bypass} ones and that the same level of $P_{\rm max}(t_h)$ is achieved 
irrespectively on the $u_0$ value. 

However, the combined analysis with the OGLE-III events clearly shows the
absence of events with expected polarization $P_{\rm max} > 1$ percent,
as a consequence of the paucity of AGB stars in the Galactic bulge.
Therefore, an observational programme devoted to measure the polarization in 
an ongoing microlensing event must be designed to be able to measure a polarization degree 
$P_{\rm max} \ut > 0.1$ percent, that is the threshold value (in 1 hour integration time)
for the FORS2 polarimeter at VLT telescope. 
Stated this, we find that, as shown in Fig. \ref{fig8},
about two dozen of events (about 1 percent of the OGLE-III events) 
have $P_{\rm max} >0.1$ percent and apparent magnitude $I<14.5$ at 
the instant $t_h$ of the polarization peak.
{
This makes realistic the possibility to measure the polarization in about a few
events per year with the present observational capabilities (OGLE-III like).
Of course, increasing the event rate (up to thousands per year) 
as expected in the forthcoming programs EUCLID, WFIRST and KMTNet, 
the chance to pick-up events with large ($P > 0.1$ percent)
polarization correspondingly increases.}

{
After the initial submission of the present manuscript
\cite{Sajadian} presented an analysis of the expected
polarization signal in the OGLE-III data set.
Although discussed within a different framework of our analysis,
their main conclusions are in agreement
with those discussed in the present work.
Specifically, a few percent of microlensing events towards
the Galactic bulge may indeed show up
detectable polarization signals.}

\section*{Acknowledgments}

{We acknowledge for stimulating discussions N. Rattenbury and H. M. Schmid.
{We also thank the anonimous referee for useful comments.}
This work make use of the IAC-Star synthetic CMD computational code.
IAC-Star is supported and maintained by the computer
division of the Instituto de Astrofisica de Canarias.
}



 

\end{document}